\documentclass[prx,aps,preprint]{revtex4-1}
\usepackage{amsmath,amssymb, graphicx}
\usepackage{bm,times, color}
\usepackage[utf8]{inputenc}
\usepackage[colorlinks]{hyperref}
\usepackage[svgnames]{xcolor}
\hypersetup{linkcolor=DarkBlue, citecolor=DarkBlue, filecolor=black, urlcolor=DarkBlue}
\usepackage{comment, mathrsfs,slashed}
\def \p{\partial}
\def \dag{\dagger}
\def \mb{\mathbf}

\def \lan{\langle}
\def \ran{\rangle}

\def \d{\mathrm{d}}

\begin{document}
\title{Universal collective modes in 2 dimensional chiral superfluids}

\author{Wei-Han Hsiao}
\affiliation{Kadanoff Center for Theoretical Physics, University of Chicago,
Chicago, Illinois 60637, USA}
\date{August 2019}

\begin{abstract}
In this work, we utilize semi-classical kinetic equations to investigate the order parameter collective modes of a class of 2 dimensional superfluids. Extending the known results for $p$-wave superfluids, we show for any chiral ground state of angular momentum $L\geq 1$, there exists a sub-gap mode with mass $\sqrt{2}\, \Delta$ in the BCS limit, where $\Delta$ is the magnitude of the ground state gap. We determine the most significant Landau parameter which contributes to the mass renormalization and show explicitly the renormalized modes become massless at the Pomeranchuk instability of the fermion vacuum. Particularly for $L=1$, we propose a continuous field theory to include the Fermi liquid effect in quadrupolar channel and produce the same result under consistent approximations. They provide potential diagnostics for distinguishing 2 dimensional chiral ground states of different angular momenta with order parameter collective modes and reveal another low energy degrees of freedom near nematic transition.    
\end{abstract}

\maketitle

\section{Introduction}
In the studies of interacting quantum many-body systems, collective modes allow physicists to explore the correlated motions of underlying degrees of freedom. Especially in a superfluid phase, the order parameter component enriches the nature of collective excitations. Paradigmatic examples include the A-  and B-phases of superfluid ${}^3$He in (3+1)D \cite{vollhardt2013superfluid}, where massive sub-gap modes exist owing to the triplet pairing structure. They manifest themselves in terms of resonant signatures of transport properties when coupled to particle-hole channel \cite{PhysRevB.62.3042}.  

These developments permit various extensions. Natural questions include: (i) Do sub-gap massive collective modes also exist in finite angular momentum pairing channels and in (2+1)D spacetime? (ii) Do these bosonic degrees of freedom acknowledge the underlying fermionic state or the property of the Fermi surface? We pay special attention to these questions mainly because of the puzzle of $\nu = \frac{5}{2}$ fractional quantum Hall state. Three of the most prominent candidates of the ground state, the Pfaffian state, $\mathsf T$-Pfaffian state, and the anti-Pfaffian state, are understood as $p+ip$, $p-ip$ and $f-if$ chiral superconductors of non-relativistic composite fermions respectively \cite{doi:10.1146/annurev-conmatphys-033117-054227}. Moreover, both experimental \cite{nphys3523, PhysRevB.96.041107, ncommu92400} and numerical \cite{PhysRevLett.121.147601} studies have revealed the importance of nematic fluctuations and quantum criticality in the second Landau level. As a consequence, a understudying of chiral superfluids / superconductors including these effects is pursued.  

Regarding (i), it is known that the 2 dimensional analog of B-phase hosts 4 modes of mass $\sqrt{2}\ \Delta$ with angular momenta $\ell = \pm 2$. $\Delta$ is the magnitude of mass of the Bogoliubov quasi-particle. Similarly, the analog of A-phase, whose fermionic spectrum is fully gapped in 2 dimensions, hosts six modes of mass $\sqrt 2\, \Delta$ \cite{BRUSOV1981,BRUSOV1982472, PhysRevLett.83.1007}. On the other hand, (ii) has been investigated in the context of (3+1)D ${}^3$He superfluid with Fermi liquid theory \cite{vollhardt2013superfluid, WOLFLE197796, SERENE1983221, PhysRevB.95.094515}, where the corrections to the masses of massive sub-gap modes and the sound speeds of Goldstone modes can be expressed in terms of Landau parameters. In addition, for Sr${}_2$RuO${}_4$ \cite{10.3389/fphy.2015.00036}, it has been shown that strong coupling effect and gap anisotropy are able to modify the magnitude of the masses and break the spectrum degeneracy. 

This work intends to address the complementary faces of (i) and (ii). We specifically focus on superfluids in (2+1)D with general pairing channels of angular momenta $L = 0, 1\cdots $. For $L= 1$, the 2 dimensional analogs of A and B phases are considered, whereas for higher $L$ we concentrate on chiral ground states. We look for massive sub-gap modes, and investigate the mechanisms that may correct their masses in long wavelength limit $q=0$.   

We find that in the limit with weak-coupling and exact particle-hole symmetry, there is at least a pair of bosonic modes of universal mass $ \sqrt{2}\, \Delta$ for all $L\geq 1$. We investigate corrections to these degenerate modes owing to fermionic vacuum in a phenomenological manner and determine the angular momentum channels substantial for mass renormalization. For a given chiral ground states of angular momentum $L$, the order parameter fluctuations longitudinal to the ground state are renormalized by the Landau parameter in the angular momentum channel $2L$, $F_{2L}$, and thus correspond to a type of spin-2$L$ mode. 

The Fermi liquid correction is especially intriguing in (2+1)D. As we will show shortly in Sec.\ref{fermiliquidcorrection}, it implies the sub-gap modes soften when $F_{2L}$ is negative. Explicitly, as $F_{2L}\to -1$ the mass of the collective modes vanishes as  
\begin{align}
\sqrt{\frac{12(1+F_{2L})}{6+F_{2L}}}\, \Delta\to 0.
\end{align}
In particular, taking $L=1$, the limit $F_{2}\to -1$ serves as one of the mechanisms behind nematic electronic phases \cite{doi:10.1146/annurev-conmatphys-070909-103925, PhysRevB.64.195109}. On top of previous studies on unconventional superconductors \cite{PhysRevLett.114.097001} and quantum Hall nematic phases \cite{nphys3523, PhysRevB.96.041107, ncommu92400}, this is another example where the Pomeranchuk instability in quadrupolar channel influences the nature of a paired phase \footnote{We note that similar indication is also found in 3 dimensional $p$-wave superfluid \cite{PhysRevB.95.094515}, but in general nematic instability is easier triggered in pure 2 spatial dimensions. An example of composite fermions is studied in a recent work Ref. \cite{2PhysRevLett.121.147601} } and it allows us to probe the high frequency spin-2 mode omitted in most literatures. We thereby propose a toy model and compute its effective action in Gaussian approximation and show the kinetic result can be captured after implementing exact particle-hole symmetry. 

Our work is new in that (i) we generalize the known high frequency sub-gap modes in $p$-wave superfluid in (2+1)D to higher angular momentum channels, and compute their mass renormalizations in terms of Landau parameters. (ii) Moreover, for $p$-wave chiral superfluids, we propose a continuous field theory model to include the Landau parameter effect in quadrupole channel. In addition to confirming the kinetic theory result, this model could easily be generalized when loosening particle-hole symmetry and provides a understanding of the underlying nature of spin-2 modes and nematic fluctuations. 

This paper is organized as follows. In Sec.~\ref{Formalism}, we review the semi-classical equation approach for the computation of collective excitations. The equations derived are used in Sec.~\ref{collectivemodes} to compute collective excitations for various ground states. In Sec.~\ref{fermiliquidcorrection}, we calculate the Fermi liquid ground state effect upon the bare bosonic spectra. Finally, Sec.~\ref{FT} presents a field theory model for a $p$-wave chiral superfluid with a continuous quadrupole interaction. We demonstrate the results in~\ref{collectivemodes} and~\ref{fermiliquidcorrection} can be produced in the limit of exact particle-hole symmetry. Finally, a summary and several open directions are composed. The full solutions to the kinetic equation~\eqref{EoM} without assuming $q=0$ and $\Delta\in\mathbb R$, and the computational method for the effective field theory are present along with the method in appendix. 


\section{Kinetic Theory}\label{Formalism}
Bosonic collective modes in superfluids or superconductors \footnote{In this work, we turn off the U(1) gauge field and therefore do not strictly distinguish these 2 terminologies.} can be computed with various approaches. In this section we start off with the time-dependent mean field approximation to include the Fermi liquid corrections. This approach can be formulated in terms of generalized Landau-Boltzmann kinetic equations \cite{vollhardt2013superfluid}, or the linearized non-equilibrium Eilenberger equation \cite{SERENE1983221, 2013arXiv1309.6018M}. Though we will not repeat the derivations of the formalism, which we refer the readers to Ref.\cite{SERENE1983221, 2013arXiv1309.6018M}, we will give a complete elaboration of the workflow. 
 
In the semi-classical limit, physical quasi-particle distribution is related to the Keldysh Green's function $\widehat{g}(\varepsilon, \hat{\mb p}; \omega, \mb q)$. In our computation it is a $4\times 4$ matrix function. We use two sets of Pauli matrices $\bm{\tau} = (\tau_1, \tau_2, \tau_3)$ and $\bm{\sigma} = (\sigma_1, \sigma_2, \sigma_3)$ to span the particle-hole space and spin space respectively. In its argument $(\varepsilon, \mb p=p_F\hat{\mb p})$ are the Fourier transformed variables of the fast coordinates, where as $(\omega, \mb q)$ are ones of the center of mass coordinates \footnote{The fast and the slow coordinates refer to the relative coordinates and the center of mass coordinates in the non-equilibrium Green's function.}. $p_F$ is the magnitude of Fermi momentum and $v_F$ is Fermi velocity. In clean limit, the linear response of a non-relativistic fermion without spin-orbital coupling is given by the following kinetic equation 
\begin{align}\label{EoM}
 \varepsilon_+\tau_3\delta\widehat{g}-\delta\widehat{g}\tau_3\varepsilon_--&v_F\hat{\mb p}\cdot\mb q\delta\widehat{g}- [\widehat{\sigma}_0, \delta\widehat{g}]\notag\\
=&\delta\widehat{\sigma}\, \widehat{g}_0(\varepsilon_-)-\widehat{g}_0(\varepsilon_+)\, \delta\widehat{\sigma} ,
\end{align}
where $\varepsilon_{\pm} $ denotes  $\varepsilon\pm \omega/2$. The operator $\widehat{\sigma}_0(\hat{\mb p})$ is the molecular mean field or the self-energy at equilibrium, while $\delta\widehat{\sigma}(\omega, \mb q)$ is the linear perturbation of $\widehat{\sigma}_0$. Similarly, $\widehat{g}_0(\varepsilon, \hat{\mb p})$ represents the Keldysh Green's function at equilibrium. It is related to retarded and advanced Green's functions via $\widehat g_0 = (g^R_0-g^A_0)\tanh (\varepsilon/2T)$, which yields \cite{2013arXiv1309.6018M}
\begin{align}
\widehat g_0 = \frac{-2\pi i(\tau_3\varepsilon-\widehat{\Delta})}{\sqrt{\varepsilon^2-|\Delta|^2}}\Theta(\varepsilon^2-|\Delta|^2)\mathrm{sgn}(\varepsilon)\tanh \frac{\varepsilon}{2T}.
\end{align}
The low energy fluctuation of quasi-particles and the deduced physical quantities are given by the $\varepsilon$-integrated $\widehat g$ 
$ \int_{-\infty}^{\infty} \frac{d\varepsilon}{2\pi i}\widehat{g}(\varepsilon, \hat{\mb p}; \omega, \mb q).$
In particular, the perturbation $\delta\widehat{\sigma}$ is self-consistently determined by the convolution of inter-particle potentials and $\delta\widehat{g}$. 

To further elaborate, we note that $\delta\widehat{g}$ has a general structure in particle-hole space
\begin{align}\label{GreenFunction}
\delta\widehat{g} = \begin{pmatrix} \delta g+\delta\mb g\cdot\bm{\sigma} & (\delta f+\delta\mb f\cdot\bm{\sigma})i\sigma_2\\ i\sigma_2(\delta f'+\delta\mb f'\cdot\bm{\sigma}) & \delta g'+\delta\mb g'\cdot\bm{\sigma}^t\end{pmatrix},
\end{align}
and accordingly so does $\delta\widehat{\sigma}$
\begin{align}\label{MolecularField}
\delta\widehat{\sigma} = \begin{pmatrix} \delta\varepsilon + \delta\bm{\varepsilon}\cdot\bm{\sigma} & (d+\mb{d}\cdot\bm{\sigma})i\sigma_2 \\ i\sigma_2( d'+\mb{d}'\cdot\bm{\sigma}) & \delta\varepsilon'+\delta\bm{\varepsilon}'\cdot\bm{\sigma}^t\end{pmatrix},
\end{align}
where the primed variables are 
\begin{subequations}
\begin{align}
&\delta g'(\hat{\mb p};\omega, \mb q) = \delta g(-\hat{\mb p}; \omega, \mb q)\\
& \delta\varepsilon'(\hat{\mb p}; \omega, \mb q) = \delta \varepsilon(-\hat{\mb p}; \omega, \mb q)\\
& \delta f'(\hat{\mb p};\omega, \mb q) = \delta f^*(\hat{\mb p}; -\omega, -\mb q)\\
& d'(\hat{\mb p}; \omega, \mb q) = d^*(\hat{\mb p}; -\omega, -\mb q).
\end{align}
\end{subequations}
We would like to explain the notations here before moving forward. The diagonal parts of $\delta\widehat{g}$ refer to the normal, or particle-hole, correlation functions $\lan \psi \psi^{\dag}\ran$ and $\lan\psi^{\dag}\psi\ran$, while the off-diagonal parts denote the anomalous, or particle-particle correlation functions $\lan\psi\psi\ran$ and $\lan\psi^{\dag}\psi^{\dag}\ran$. $\bm{\sigma}$ denotes the Pauli matrices $(\sigma_1, \sigma_2, \sigma_3)$ in spin space as defined earlier and $\bm{\sigma}^t$ denotes the transposed Pauli matrices. Looking at the Green's function $\delta\widehat g$, in the particle-hole channel, $\delta g$ and $\delta \mb g$ denote spin-independent and spin-dependent correlations respectively. In the particle-particle channel, $\delta f$ represent the spin-singlet pairing amplitude and $\delta \mb f$ the spin-triplet one. Correspondingly, the diagonal part of the self-energy $\delta\widehat{\sigma}$ is the particle-hole self-energy, including the spin-independent $\delta\varepsilon$ and spin-dependent part $\delta\bm{\varepsilon}$. The off-diagonal part of the self-energy is the superfluid gap induced by anomalous correlations. $d$ is the spin-singlet gap and $\mb d$ denotes the spin-triplet gap.

Note that physical observables are usually expressed in terms of the symmetric and anti-symmetric combination of $\delta g$, $\delta f$ and their primed partners. In this work, we define $(+)$ and $(-)$ combinations of a function $f$ as
\begin{align}\label{def_pm}
f^{(\pm)} = f\pm f'.
\end{align}
The eigenvalues $(\pm 1)$ represent the parity under charge conjugation. As we will see, the charge density and energy stress tensor correspond to the scalar and quadrupole modes of $\delta g^{(+)}$ respectively, whereas the current density is proportional to the vector mode of $\delta g^{(-)}$. Similarly, $\delta f^{(+)}$ and $\delta f^{(-)}$ stand for the amplitude and phase fluctuations of the anomalous correlation functions. 

 To complete the equations, the correction to the self-energy is determined by the two-body vertex. Evaluating internal momentum integral over the Fermi surface, we have, in the particle-hole channel \cite{fermiliquid, abrikosov1975methods, landau1980statistical}, 
\begin{subequations}
\begin{align}
\label{F1}\delta\varepsilon(\hat{\mb p} ;  \omega, \mb q ) = \delta\varepsilon_{\rm ext}(\hat{\mb p}; \omega, \mb q)+\int \frac{d\theta'}{2\pi}A^s(\theta, \theta')\int \frac{d\varepsilon'}{4\pi i} \delta {g}(\varepsilon', \hat{\mb p}'; \omega, \mb q),
\end{align}
\begin{align}
\label{F2}\delta\bm{\varepsilon}(\hat{\mb p}; \omega, \mb q  ) = \delta\bm{\varepsilon}_{\rm ext}(\hat{\mb p}; \omega, \mb q)+\int \frac{d\theta'}{2\pi}A^a(\theta, \theta')\int \frac{d\varepsilon'}{4\pi i} \delta \mb{g}(\varepsilon', \hat{\mb p}'; \omega, \mb q).
\end{align}
\end{subequations}
These 2 equations state that at 1-loop the particle-hole self-energy consists of external perturbation $\delta\varepsilon_{\rm ext}$ or $\delta\bm{\varepsilon}_{\rm ext}$ and a fermion loop closed by a two-body interaction vertex. An example of $\delta\varepsilon_{\rm ext}$ is a background inhomogeneous chemical potential, whereas an example of $\delta\bm{\varepsilon}_{\rm ext}$ could be a weak external magnetic field. $A^s$ $(A^a)$ is the spin-independent (exchange) forward scattering amplitude which can be rewritten in terms of Landau parameters $F$ via the relation
\begin{align}\label{A_in_F}
A(\theta, \theta') =F(\theta, \theta') -\int\frac{d\theta''}{2\pi}F(\theta, \theta'')A(\theta'', \theta'). 
\end{align}
Similar expressions arise in the particle-particle channel. Since the fluctuations of the superfluid gaps directly come from the anomalous correlation functions, the off-diagonal components are related by the linearized gap equations.
\begin{subequations}
\begin{align}\label{G1}
& d(\hat{\mb p}; \omega, \mb q) = \int\frac{d\theta'}{2\pi}V_{\rm e}(\theta, \theta')\int \frac{d\varepsilon'}{4\pi i}\delta f(\varepsilon', \hat{\mb p}'; \omega, \mb q)
\end{align}
\begin{align}\label{G2}
&\mb d(\hat{\mb p}; \omega, \mb q) = \int \frac{d\theta'}{2\pi}V_{\rm o}(\theta, \theta')\int \frac{d\varepsilon'}{4\pi i}\delta\mb f(\varepsilon', \hat{\mb p}'; \omega, \mb q),
\end{align}
\end{subequations}
where $V_{\rm e}$ ($V_{\rm o}$) is the pairing potentials in even (odd) angular momentum channel.

In 2 dimensions, the scattering amplitudes and pairing potentials yield the approximate angular expansions
\begin{subequations}
\begin{align}
& A = \sum_{\ell = -\infty}^{\infty}A_{\ell} e^{-i\ell(\theta-\theta')},\ A_{\ell} = A_{-\ell},\\
& V_{\rm e} = \sum_{\ell\in\{\rm even\}}V_{\ell}[e^{-i\ell(\theta-\theta')}+\mathrm{h.c.}]\\
& V_{\rm o} = \sum_{\ell\in\{\rm odd\}}V_{\ell}[e^{-i\ell(\theta-\theta')}+\mathrm{h.c.}].
\end{align}
\end{subequations}
from which and~\eqref{A_in_F} we can derive $A_{\ell} = \frac{F_{\ell}}{1+F_{\ell}}$, where $F_{\ell}$ is the conventional dimensionless Landau parameter of angular momentum channel $\ell$. Notation-wise, for other functions $f(\hat{\mb p})$ evaluated at a point on the Fermi surface $\hat{\mb p}$, the angular decomposition is defined as
\begin{align}\label{angular_Fourier}
f = \sum_{\ell = -\infty}^{\infty}e^{-i\ell\theta}f_{\ell}.
\end{align}
We can then provide a recipe for the computation. We first invert~\eqref{EoM} to obtain the perturbed Green's function $\delta\widehat g$ as a function of equilibrium Green's function $\widehat g_0$, equilibrium self-energy $\widehat{\sigma}_0$ and perturbed self-energy $\delta\widehat{\sigma}$. Taking the convolution as in~\eqref{F1},~\eqref{F2},~\eqref{G1}, and~\eqref{G2} establishes integral equations for $\delta\widehat{\sigma}$. Projecting equations~\eqref{G1} and~\eqref{G2} to different angular modes $\ell$ gives us the coupled equations of $d_{\ell}$, $\mb{d}_{\ell}$, $\delta\varepsilon$, and $\delta\bm{\varepsilon}$. The bare bosonic collective modes are given by the normal modes of the homogeneous part of the equations. To include the Fermi liquid corrections, we project~\eqref{F1}, and~\eqref{F2} to their $\ell$th angular modes as well and solve $\delta\varepsilon_{\ell}$ and $\delta\bm{\varepsilon}_{\ell}$ in terms of $\delta\varepsilon_{\rm ext}$, $\delta\bm{\varepsilon}_{\rm ext}$ $d$ and $\mb d$. Plugging the results back into the equations for $d_{\ell}$ and $\mb d_{\ell}$ yields inhomogeneous equations sourced solely by external fields. The renormalized mass spectrum is solved as the poles of the solution kernels. 

In the rest of this section, we use the above formulation to derive the integral equation for 2 dimensional spin-singlet and spin-triplet superfluids and compute the collective modes and Fermi liquid corrections in the sections following. While in the main text only the equations in long wavelength limit are presented, the complete set of dynamical equations are given in appendix \ref{FDE}.  
\subsection{Spin-singlet pairing}
In a spin-singlet pairing channel, the equilibrium self-energy is characterized by a complex gap field $\Delta$. 
\begin{align}\label{singletgap}
\widehat{\sigma}_0 = \widehat{\Delta} = \begin{pmatrix} 0 & \Delta i\sigma_2 \\ \Delta^* i\sigma_2 & 0 \end{pmatrix}.
\end{align}
The fluctuation of the spin-singlet order parameter can be parametrized by a complex number $d$. It transforms as a scalar under spin rotation SO${}_S$(3) and can have internal structures, i.e., tensor indices under orbital rotation SO${}_L$(2) depending on pairing symmetries. In the absence of magnetic field, spin-triplet fluctuations $\mb d$ decouple from $d$. Hence we consider them separately in the present work. 

Plugging~\eqref{singletgap} into~\eqref{EoM}, inverting it using the variables defined in~\eqref{GreenFunction} and~\eqref{MolecularField}, and taking the convolution as in~\eqref{G1} and~\eqref{G2} give us, in the long-wavelength limit, the off-diagonal components of the molecular fields  
\begin{subequations}
\begin{align}\label{singletd1}
d(\hat{\mb p}; \omega) = \int \frac{d\theta'}{2\pi}V_{\rm e}(\theta, \theta')&\Big[\Big(\gamma+\frac{1}{4}\bar{\lambda}[\omega^2-2|\Delta|^2]\Big)d-\frac{\bar{\lambda}}{2}\Delta^2d'-\frac{\omega}{4}\bar{\lambda}\Delta\delta\varepsilon^{(+)}\Big],
\end{align}
\begin{align}\label{singletd2}
d'(\hat{\mb p}; \omega) = \int\frac{d\theta'}{2\pi}&V_{\rm e}(\theta, \theta')  \Big[\Big(\gamma+\frac{1}{4}\bar{\lambda}[\omega^2-2|\Delta|^2]\Big)d' -\frac{\bar{\lambda}}{2}(\Delta^*)^2d+\frac{\omega\bar{\lambda}}{4}\Delta^*\delta\varepsilon^{(+)}\Big].
\end{align}
\end{subequations}
$\gamma$ is the BCS logarithm given explicitly in Appendix~\eqref{gammafunction}. The function $\lambda$, often called the Tsunedo function, whose complete form is given in appendix~\ref{Fxns}. In $q\to 0$ limit, 
\begin{align}
\bar{\lambda} =\frac{\lambda(\hat{\mb p};\omega)}{|\Delta|^2} = \int_{|\Delta|}^{\infty}\frac{d\varepsilon}{\sqrt{\varepsilon^2-|\Delta|^2}}\frac{\tanh\frac{\varepsilon}{2T}}{\varepsilon^2-\omega^2/4}.
\end{align}
There could be angular dependence through the anisotropy in $|\Delta|^2$ even in the long wavelength limit. Suppose only a single pairing channel $L$ is significant, i.e., that $V = V_L(e^{-iL(\theta-\theta')}+\rm h.c.)$. Taking $\int \frac{d\theta}{2\pi} e^{iL\theta}$ on both sides of~\eqref{singletd1} and~\eqref{singletd2} eliminates $\gamma$s. 
The dynamical equations of motion are then obtained 
\begin{subequations}
\begin{align}
\label{dyneqn1}& \Big\lan e^{iL\theta}\bar{\lambda}\Big([\omega^2-2|\Delta|^2]d-2\Delta^2d'-\omega\Delta\delta\varepsilon^{(+)}\Big)\Big\ran =0\\
\label{dyneqn2}& \Big\lan e^{iL\theta}\bar{\lambda}\Big([\omega^2-2|\Delta|^2]d'-2(\Delta^*)^2d+\omega\Delta^*\delta\varepsilon^{(+)}\Big)\Big\ran =0,
\end{align}
\end{subequations}
where we use the angle bracket $\lan \cdots\ran$ to denote the angular average $\int_{-\pi}^{\pi} \frac{d\theta}{2\pi}\cdots$.
\subsection{spin-triplet pairing}
In a spin-triplet pairing channel, the ground state self-energy is characterized by the vector-valued gap function $\bm{\Delta}$ 
\begin{align}
\widehat{\Delta} = \begin{pmatrix} 0 & \bm{\Delta} \cdot i\bm{\sigma}\sigma_2 \\ \bm{\Delta}^*\cdot i\sigma_2\bm{\sigma} & 0 \end{pmatrix}.
\end{align}
The fluctuation is encoded in the dynamics of the $\mb d$ vector, which transforms as a vector under SO${}_S$(3), and could contain internal structure depending on pairing symmetry as well. Taking 2 dimensional $p$-wave superfluids for example, it can be expanded as $d_{\mu}(\hat{\mb p}) = d_{\mu i}\hat p_i$, where $i = x, y$. Inverting the kinetic equations, the dynamical equations for $\mb d$ in $q\to 0$ limit are 

\begin{subequations}
\begin{align}\label{tripletd1}
\mb d = & \int \frac{d\theta'}{2\pi}V_{\rm o}(\theta,\theta')\Big[\Big(\gamma+\frac{1}{4}\bar{\lambda}(\omega^2-2|\Delta|^2)\Big)\mb d\notag\\
& +\frac{\bar{\lambda}}{2}[(\bm{\Delta}\cdot\bm{\Delta})\mb d'-2(\bm{\Delta}\cdot\mb{d}')\bm{\Delta}]-\frac{\omega\bar{\lambda}}{4}(\mb{\Delta}\delta\varepsilon^{(+)}-i\bm{\Delta}\times\delta\bm{\varepsilon}^{(+)})\Big].
\end{align}
\begin{align}\label{tripletd2}
\mb d' = & \int \frac{d\theta'}{2\pi} V_{\rm o}(\theta, \theta')\Big[\Big(\gamma+\frac{1}{4}\bar{\lambda}(\omega^2-2|\Delta|^2)\Big)\mb d'\notag\\
& +\frac{\bar{\lambda}}{2}[(\bm{\Delta}^*\cdot\bm{\Delta}^*)\mb d-2(\bm{\Delta}^*\cdot\mb d)\bm{\Delta}^*]+\frac{\omega\bar{\lambda}}{4}(\bm{\Delta}^*\delta\varepsilon^{(+)}+i\bm{\Delta}^*\times\delta\bm{\varepsilon}^{(+)}).
\end{align}
\end{subequations}
Again we multiply~\eqref{tripletd1}, and~\eqref{tripletd2} by $V = V_L[e^{-iL(\theta-\theta')}+e^{iL(\theta-\theta')}]$ and integrate over $\theta$, giving us \begin{subequations}
\begin{align}\label{dyneqn3}
&\Big\lan e^{iL\theta}\bar{\lambda}\Big( [\omega^2-2|\Delta|^2]\mb d+2(\bm{\Delta}\cdot\bm{\Delta})\mb d'-4(\bm{\Delta}\cdot\mb d')\bm{\Delta}\Big)\Big\ran\notag\\
= & \omega\Big\lan e^{iL\theta}\bar{\lambda}(\bm{\Delta}\delta\varepsilon^{(+)}-i\bm{\Delta}\times\delta\bm{\varepsilon}^{(+)})\Big\ran.
\end{align}
\begin{align}\label{dyneqn4}
& \Big\lan e^{i L\theta}\bar{\lambda}\Big([\omega^2-2|\Delta|^2]\mb d'+2(\bm{\Delta}^*\cdot\bm{\Delta}^*)\mb d-4(\bm{\Delta}^*\cdot\mb d)\bm{\Delta}^*\Big)\Big\ran\notag\\
= & -\omega\Big\lan e^{iL\theta}\bar{\lambda}(\bm{\Delta}^*\delta\varepsilon^{(+)}+i\bm{\Delta}^*\times\delta\bm{\varepsilon}^{(+)})\Big\ran.
\end{align}
\end{subequations}
In the following section, we will solve~\eqref{dyneqn1},~\eqref{dyneqn2},~\eqref{dyneqn3}, and~\eqref{dyneqn4} for ground states of different pairing channels and symmetries.

\section{Collective Modes}\label{collectivemodes}
In this section we utilize the equations derived in the last section to compute the bare bosonic spectra for various superconducting ground states. 
We focus on chiral ground states of angular momentum $L\neq  0$, in which the massive collective modes are interpreted as spin-$2L$ modes. The masses of order parameter collective modes appear as normal modes of the homogeneous part in ~\eqref{dyneqn1},~\eqref{dyneqn2},~\eqref{dyneqn3}, and~\eqref{dyneqn4}. The self-energy $\delta\varepsilon$ and $\delta\bm{\varepsilon}$ in Landau channel are treated as external sources at the zeroth order, and they will {\it renormalize} the above bare masses in the next section as we conclude Fermi liquid effects. 
\subsection{$s$-wave pairing}
For $s$-wave pairing, it is possible to choose a gauge such that $\Delta\in\mathbb R$, in which limit the amplitude mode $d^{(+)}$ and phase mode $d^{(-)}$ decouple. The superscripts $(+)$ and $(-)$ are defined according to~\eqref{def_pm}. The bosonic field has no internal structure and is simply a complex scalar. Two order parameter collective modes thus exist and obey the equations
\begin{subequations}
\begin{align}
\label{swave1}& (\omega^2-4\Delta^2)d^{(+)} = 0\\
\label{swave2}& \omega^2d^{(-)} = 2\omega\Delta \delta\varepsilon^{(+)}_0,
\end{align}
 \end{subequations}
 where the zero-angular momentum quasi-particle energy $\delta\varepsilon^{(+)}_0$ is obtained under the projection~\eqref{angular_Fourier}. The normal modes have masses $2\Delta$ and $0$ corresponding to the simplest example of {\it Higgs} and {\it Goldstone} bosons respectively. Note that if we compute~\eqref{swave2} to the leading non-vanishing order in $q^2$, we would have obtain $(\omega^2-\frac{1}{2}(v_Fq)^2)d^{(-)}$, entailing the Goldstone boson moves at the speed $v_F/\sqrt 2$. Another observation is that the Higgs mode receives no external force and consequently it would not be renormalized by particle-hole self-energy. On the other hand, the Goldstone boson is sourced by the density mode $\delta\varepsilon^{(+)}_0$, which would trigger Higgs mechanism in the presence of Coulomb interaction.  
\subsection{$p$-wave pairing}
Owing to triplet-pairing and orbital structure, the p-wave pairing states have more degrees of freedom and thus more collective modes. In 2 dimensions, the fluctuation of $p$-wave superconductors can be represented by the complex tensor $d_{\mu i}$, which contains $3\times 2$ complex degrees of freedom, leading to 12 collective modes in total. The number of the massless modes $N_G$, as we will see shortly, can be determined by ground state symmetry breaking pattern. The rest $(6-N_G)\times 2$ is number of sub-gap collective modes.   
\paragraph{B-phase}
We first consider the 2-dimensional analog of ${}^3$He B-phase, where the gap function assumes the form
\begin{align}\label{bphasegap}
\bm{\Delta} = \frac{\Delta}{p_F}(\hat{\mb x}p_x+\hat{\mb y}p_y),\ \Delta\in\mathbb R.
\end{align} 
In this phase, the global symmetry breaks following the pattern SO${}_S$(3)$\otimes$SO${}_L$(2)$\otimes$U(1)$\to$ SO(2), which immediately indicates the existence of 4 Goldstone modes. Besides, the residual symmetry is SO${}$(2) rotation and we expect the fluctuations can be characterized by total angular momentum $J$. 
Owing to this fact, it is convenient to first decompose $d_{\mu}$ into different angular momentum channels $d_{\mu} = \sum_{m = \pm 1}d_{\mu m}e^{-im\theta}$, where $\theta$ is the polar angle of $\hat{\mb p}$, and take the linear combinations as follows 
\begin{align}
& D_{\pm m} = d_{x m}\pm i d_{ym}\\
& D_{0 m} = d_{zm}.
\end{align}
These $D_{\sigma\sigma'}$'s form a nice basis in which the dynamical equations can be solved. 
Moreover, as the gap function is real, modes transforming differently under charge conjugation again decouple. That is to say, we can further separate $d^{(\pm)} = d\pm d'$ degrees of freedom. We first look at the $\mb d^{(-)}$ modes governed by the equation
\begin{align}
(\omega^2-4\Delta^2)\mb d^{(-)}+4(\mb{\Delta}\cdot\mb d^{(-)})\mb{\Delta} = 2\omega\mb{\Delta}\delta\varepsilon^{(+)}.
\end{align}
Organizing the dynamical equations using the basis $D_{\sigma\sigma'}$, we could find
\begin{subequations}
\begin{align}
& (\omega^2-4\Delta^2)D_{0\pm}^{(-)} = 0\\
\label{bphaseminus2}& (\omega^2-2\Delta^2)D_{\pm\pm}^{(-)} = 2\omega\Delta \delta\varepsilon^{(+)}_{\pm 2}\\
\label{bphaseminus3}& (\omega^2-4\Delta^2)(D^{(-)}_{+-}-D^{(-)}_{+-})= 0\\
& \omega^2(D_{+-}^{(-)}+D_{-+}^{(-)})= 4\omega\Delta\delta\varepsilon^{(+)}_0.
\end{align}
\end{subequations}
Consequently, $d^{(-)}$ has 2 sub-gap massive modes $J= \pm 2$ of the same mass $\sqrt 2\, \Delta$, sourced by the spin-independent quadrupolar molecular field $\delta\varepsilon^{(+)}_{\pm 2}$. 

Next we look at $d^{(+)}$, which obeys
\begin{align}
[\omega^2\mb d^{(+)}-4\bm{\Delta}(\bm{\Delta}\cdot\mb d^{(+)})] = -2i\omega\bm{\Delta}\times\delta\bm{\varepsilon}^{(+)}.
\end{align}
Following the same procedure to project each component to different $J$ sectors, we would obtain 
\begin{subequations}
\begin{align}
\label{bphaseplus1}& \omega^2 d_{0}^{(+)}=-2i\omega (\bm{\Delta}\times\delta\bm{\varepsilon}^{(+)})\cdot\hat{\mb z}\\
\label{bphaseplus2}& (\omega^2-2\Delta^2)D_{\pm\pm}^{(+)}= \mp 2\omega\Delta\delta\bm{\varepsilon}^{(+)}_{\pm 2}\cdot\hat{\mb z}\\
\label{bphaseplus3}& \omega^2 (D_{-+}^{(+)}-D_{+-}^{(+)}) = 4\Delta\omega\delta\bm{\varepsilon}^{(+)}_0\cdot\hat{\mb z}\\
& (\omega^2-4\Delta^2)(D^{(+)}_{+-}+D^{(+)}_{+-}) = 0.
\end{align}
\end{subequations}
Again modes $D_{\pm\pm}^{(+)}$ have rest mass $\sqrt{2}\, \Delta$ and they are driven by the $z$-component of the spin-dependent quadrupolar fields $\delta\bm{\varepsilon}_{\pm 2}^{(+)}\cdot\hat{\mb z}$.

\paragraph{A-phase} Considering only the continuous symmetry, 2 dimensional A-phase has a different symmetry breaking pattern SO${}_S$(3)$\otimes$SO${}_L$(2)$\otimes$U(1)$\to$ U${}_{L-N/2}$(1)$\otimes$U${}_{S_z}$(1). The residual symmetry contains 2 parts. U${}_{L-N/2}$(1) refers to the combination of orbital and phase rotation. The order parameter is symmetric when an orbital rotation of angle $\alpha$ is followed by a phase rotation $-\alpha/2$. The U${}_{S_z}$(1) is the residual spin rotation about the direction of ground state $\bm{\Delta}$.

Let us now consider a $p+ip$ ground state described by 
\begin{align}\label{pwavegap}
\bm{\Delta} =\frac{p_x+ip_y}{p_F}\Delta\hat{\mb z}= e^{i\theta}\Delta \hat{\mb z},\ \Delta\in\mathbb R. 
\end{align}
The dynamic equations for $\mb d_{\ell}$ and $\mb d'_{\ell}$ ($\ell = \pm 1$) are now coupled and given as follows. 
\begin{subequations}
\begin{align}\label{chiralEqn1}
&\lan e^{i\ell\theta}[(\omega^2-2\Delta^2)\mb d+2\Delta^2\mb d' e^{2i\theta}-4\Delta^2\hat{\mb z}(\hat{\mb z}\cdot\mb d')e^{i2\theta}]\ran\notag\\
= & \omega\Delta\lan e^{i\ell\theta}e^{i\theta}[\hat{\mb z}\delta\varepsilon^{(+)}-i\hat{\mb z}\times\delta\bm{\varepsilon}^{(+)}]\ran.
\end{align}
\begin{align}\label{chiralEqn2}
& \lan e^{i\ell\theta}[(\omega^2-2\Delta^2)\mb d'+2\Delta^2 \mb de^{-2i\theta}-4\Delta^2\hat{\mb z}(\hat{\mb z}\cdot\mb d)e^{-i2\theta}]\ran\notag\\
= & -\omega\Delta\lan e^{i\ell\theta}e^{-i\theta}[\hat{\mb z}\delta\varepsilon^{(+)}+i\hat{\mb z}\times\delta\bm{\varepsilon}^{(+)}]\ran.
\end{align}
\end{subequations}
We first look at the angular modes $\mb d_{\ell =1}$ and $\mb d'_{\ell =-1}$. They obey the equations 
\begin{subequations}
\begin{align}
\label{chiralp1}& (\omega^2-2\Delta^2)\mb d_1 = \omega\Delta\lan\hat{\mb z}\delta\varepsilon^{(+)}_2-i\hat{\mb z}\times\delta\bm{\varepsilon}^{(+)}_2 \ran\\
\label{chiralp2}& (\omega^2-2\Delta^2)\mb d_{-1}' = -\omega\Delta\lan \hat{\mb z}\delta\varepsilon^{(+)}_{-2}+i\hat{\mb z}\times\delta\bm{\varepsilon}^{(+)}_{-2}\ran 
\end{align} 
\end{subequations}
and have the same mass $\sqrt 2\, \Delta$. The external forces consist of both spin-dependent and spin-independent molecular fields, both of which are projected to quadrupolar channels. On the other hand, equations for $\mb d_{\ell=-1}$ and $\mb d'_{\ell=1}$ are coupled. Solving these equations, one can find 3 massless modes and 3 modes of mass $2\Delta$. The external forces on the right-hand sides of~\eqref{chiralp1} and~\eqref{chiralp2} consist of both spin-dependent and spin-independent molecular fields, both of which are projected to quadrupolar channels. 

\subsection{$d$-wave pairing}
The $d$-wave gap fluctuation is captured by the complex field $d_{ij}\hat p_i\hat p_j$ with irreducible complex degrees of freedom $1\times 2$, represented by the modes $d_{\pm 2}e^{\mp i2\theta}.$ In this work, we consider the chiral ground state 
\begin{align}
\frac{\Delta}{p_F^2} ( p_x+i p_y)^2=\Delta e^{i2\theta},\ \Delta\in\mathbb R.
\end{align}
Equations~\eqref{dyneqn1}, and~\eqref{dyneqn2} then become 
\begin{subequations}
\begin{align}
\label{chiralEqn3}& (\omega^2-2\Delta^2)d_2 = \omega\Delta\delta\varepsilon^{(+)}_4\\
\label{chiralEqn4}& (\omega^2-2\Delta^2)d_{-2}' = -\omega\Delta\varepsilon^{(+)}_{-4}\\
& (\omega^2-4\Delta^2)(d_{-2}+d_{2}') = 0\\
& \omega^2 (d_{-2}-d'_2) = 2\omega\Delta\delta\varepsilon^{(+)}_0.
\end{align}
\end{subequations}
Clearly $d_2$ and $d_{-2}'$ have masses $\sqrt{2}\, \Delta$ and the external driving forces have angular momenta $\pm 4$. 
\subsection{Higher $L$ chiral ground states}
Extending the analyses for $p$- and $d$-channels, we could actually consider a more general ground state 
\begin{subequations}
\begin{align}
& \mathrm{singlet}: \Delta e^{iL_s\theta},\ L_s = \rm even\\
& \mathrm{triplet}: \hat{\mb z}\Delta e^{iL_t\theta},\ L_t = \rm odd.
\end{align}
\end{subequations}
Modes $d_{L_s}$, $d'_{-L_s}$, $\mb d_{L_t}$ and $\mb d'_{-L_t}$ would automatically satisfy 
\begin{subequations}
\begin{align}
& (\omega^2-2\Delta^2)d_{L_s} = \omega\Delta\delta\varepsilon^{(+)}_{2L_s}\\
& (\omega^2-2\Delta^2)d_{-L_s} = -\omega \Delta \delta\varepsilon^{(+)}_{-2L_s}\\
& (\omega^2-2\Delta^2)\hat{\mb z}\cdot\mb d_{L_t} = \omega\Delta \delta\varepsilon^{(+)}_{2L_t}\\
&(\omega^2-2\Delta^2)\hat{\mb z}\cdot\mb d_{-L_t}'= -\omega\Delta\delta\varepsilon^{(+)}_{-2L_t}.
\end{align}
\end{subequations}
In this sense, $\sqrt 2\Delta$ is a universal order parameter collective mode for any chiral ground state of angular momentum $L$, each of which is sourced by quasi-particle self-energy $\delta \varepsilon_{2L}$. Since the right-hand sides belong to specific angular momentum channels, the collective modes could be regarded as generalized spin-$2L$ modes. 

\section{Fermi Liquid Corrections}\label{fermiliquidcorrection}
In the previous section we found for chiral ground states of given $L$, spin-$2L$ bosonic modes $d_L$ and $d'_{-L}$ have finite mass $\sqrt 2\Delta$. In this section we compute the Fermi liquid corrections to the mass spectra. Before presenting quantitative details, we point out some general features. Those modes with mass $ 2\Delta$, e.g. equation~\eqref{swave1}, in general are not sourced by fermionic self-energy, and consequently these modes are not renormalized. On the other hand, for those massless modes, e.g. equation~\eqref{swave2}, short-range fermionic self-energy can at most renormalize the sound speed and the magnitude of external source fields instead of generating a gap. We will therefore focus on the spin-$2L$ modes of mass $\sqrt 2\, \Delta$. 
\subsection{Massless Modes}
Let us first look at the massless modes in the s-wave channel~\eqref{swave2}. The right hand side $\delta\varepsilon^{(+)}_0$ consists of pure external perturbation and the renormalization coming form the integral part of~\eqref{F1}. Since we have rewritten the equation~\eqref{F1} and~\eqref{F2} using $F(\theta, \theta')$ instead of $A(\theta, \theta')$, we substitute the properly normalized external perturbations $\delta\varepsilon_{\rm ext}$ and $\delta\bm{\varepsilon}_{\rm ext}$ with new symbols $\delta u$ and $\delta\mb u$. In long wavelength limit, 
\begin{align}
&\delta\varepsilon^{(+)}(\theta)= \delta u^{(+)}+\int \frac{d\theta'}{2\pi}F^s(\theta, \theta')[-\lambda\delta\varepsilon^{(+)}+\frac{\omega\lambda}{2\Delta}d^{(-)}].
\end{align}
Projecting out $\ell = 0$ component, we obtain 
\begin{align}
(1+\lambda(\omega)F_0^s)\delta\varepsilon^{(+)}_0 = \delta u_0^{(+)}+\frac{\omega\lambda}{2\Delta}F_0d^{(-)},
\end{align}
plugging which back into~\eqref{swave2} yields 
\begin{align}
\omega^2d^{(-)} =2\omega\Delta \delta u_0.
\end{align}
It entails that $d^{(-)}$ remains massless. To demonstrate a triplet-pairing example, we look at B-phase~\eqref{bphasegap} and~\eqref{bphaseminus3}. For triplet-pairing states, the diagonal term of~\eqref{F1} reads
\begin{align}
& \delta\varepsilon^{(+)}(\theta) = \delta u^{(+)}+\int \frac{d\theta'}{2\pi}F^s(\theta, \theta')\Big[-\lambda(\omega)\delta\varepsilon^{(+)}+\frac{1}{2}\omega\bar{\lambda}\bm{\Delta}\cdot\mb d^{(-)}\Big],
\end{align}
whose projection to $\ell$th mode is 
\begin{align}
\label{bphaseselfE}\delta\varepsilon^{(+)}_{\ell} = \frac{\delta u^{(+)}_{\ell}+\frac{1}{2}\bar{\lambda}\omega F^s_{\ell}(\bm{\Delta}\cdot\mb{d}^{(-)})_{\ell}}{1+\lambda(\omega)F^s_{\ell}}.
\end{align}
For $\ell = 0$,
\begin{align}
\mathrm B: \delta\varepsilon_0^{(+)} = \frac{\delta u_0+\frac{\lambda\omega}{4\Delta}F_0^s(D_{+-}^{(-)}+D_{-+}^{(-)})}{1+\lambda F_0^s}
\end{align}
and we again find 
\begin{align}
\omega^2 (D_{+-}^{(-)}+D_{-+}^{(-)}) =4\omega\Delta\delta u_0.
\end{align}
The dynamical equations for $d_0$ and $D_{\{+-\}}$ are not modified by $F_0^s$, which implies a short-range interaction is not capable of gapping the Goldstone mode. 

\subsection{Massive Sub-gap Modes}
Let us continue to examine how Landau parameters renormalize massive modes. We start with the B-phase~\eqref{bphaseminus2}. Take $\ell=\pm 2$ component of~\eqref{bphaseselfE}. 
\begin{align}
\mathrm{B}: \delta\varepsilon^{(+)}_{\pm 2} = \frac{\delta u_{\pm 2}+\frac{\omega\lambda}{4\Delta}F_2^sD^{(-)}_{\pm\pm}}{1+\lambda F_2^s}.
\end{align}
Plugging this back into~\eqref{bphaseminus2} renormalizes the solutions as 
\begin{align}
D_{\pm\pm}^{(-)}= \frac{2\omega\Delta\delta u_{\pm2}}{(\omega^2-2\Delta^2)+\frac{1}{2}\lambda F_2^s(\omega^2-4\Delta^2)}.
\end{align}
The new mass is given by the zero of the denominator. In the limit $|F_2^s|\ll1$, 
\begin{align}
\omega^2\simeq 2\Delta^2(1+\textstyle \frac{1}{2}\lambda F_2^s).
\end{align}
$\lambda$ is a positive number of order 1. We can see modes get heavier for repulsive interactions $F_2^s>0$ and soften for attractive interactions $F_2^s<0$.  

Next let us look at the mode in the A-phase~\eqref{bphaseplus2} sourced by spin-dependent quasi-particle energy. 
\begin{align}
\delta&\varepsilon^{(+)}_z (\theta)= \delta u_z+\int \frac{d\theta'}{2\pi}F^a(\theta, \theta') \Big[-\lambda\delta\varepsilon^{(+)}_z-\frac{i\omega}{2}\bar{\lambda}(\bm{\Delta}\times\mb d^{(+)})_z\Big]
\end{align}
with $\delta u_z = \delta \mb u\cdot\hat{\mb z}$. Projecting it to $\ell = \pm2$ modes, 
\begin{align}
\delta\varepsilon^{(+)}_{z,\pm2} = \frac{\delta u_{z,\pm 2}\mp F_2^a\frac{\omega\lambda}{4\Delta}D_{\pm\pm}^{(+)}}{1+\lambda F_2^a}.
\end{align}
Substituting this back into~\eqref{bphaseplus2} yields 
\begin{align}
D_{\pm\pm}^{(+)} = \frac{\pm 2\omega\Delta\delta u_{z,{\pm2}}}{(\omega^2-2\Delta^2)+\frac{1}{2}\lambda F_2^a(\omega^2-4\Delta^2)}.
\end{align}
Therefore, the mass correction is given by the same transcendental equation with the replacement $F_2^s\to F_2^a$.

We are now ready to repeat the above computation for general chiral ground states. As it can be inferred from the previous analyses, the equations for singlet-pairing states are identical to ones for the longitudinal components $(\mb d\cdot\bm{\Delta})$  of the triplet-pairing states. Moreover, higher $L$ states also have the same algebraic forms. Hence, we will concentrate on triplet-pairing states and take $L=1$ without loss of generality. 

The main difference between the preceding analyses and the one for general chiral states is that the gap function can no longer be chosen real and $d^{(\pm)}$ are no longer a good basis. Consequently, the scalar self-energy would satisfy the equation 
\begin{align}\label{chiralSelfEnergy}
\delta\varepsilon^{(+)} = \delta u^++\int \frac{d\theta'}{2\pi}&F(\theta, \theta') \Big[-\lambda \delta\varepsilon^{(+)} +\frac{1}{2}\omega\bar{\lambda}(\bm{\Delta}\cdot\mb d-\bm{\Delta}\cdot\mb d')\Big].
\end{align}

\begin{figure}
\includegraphics[width = 0.8\columnwidth]{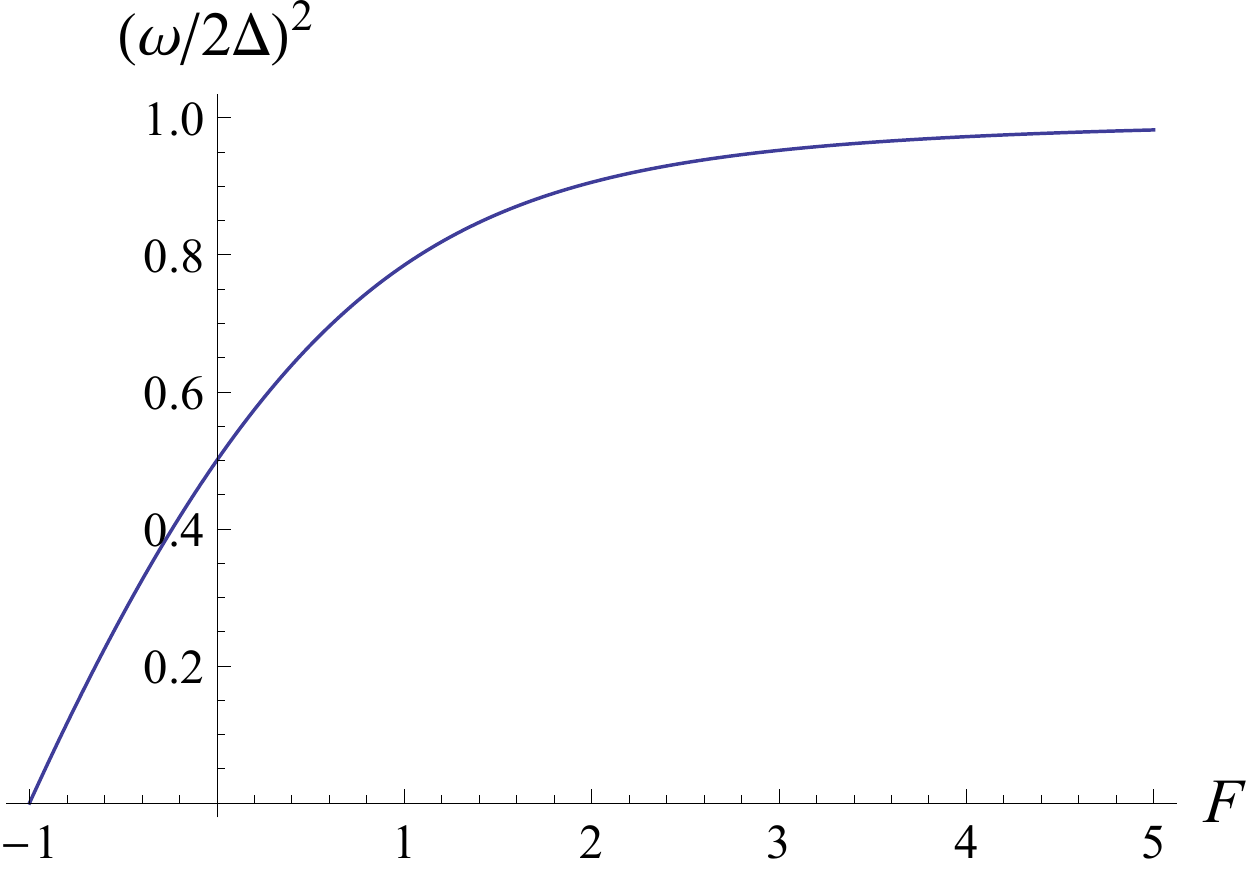}
\caption{\label{fig1} The root to~\eqref{MassEqn} depending on the value of Landau parameter $F$.}
\end{figure}

Let us take the $z$ component of~\eqref{chiralp1} and~\eqref{chiralp2}
\begin{align}
& (\omega^2-2\Delta^2)d_{1z} = \omega\Delta \delta\varepsilon^{(+)}_2\\
& (\omega^2-2\Delta^2)d_{-1z}' = -\omega\Delta \delta\varepsilon^{(+)}_{-2}.
\end{align}
Renormalizing $\delta\varepsilon^{(+)}_{\pm 2}$ with~\eqref{chiralSelfEnergy}, we find
\begin{subequations} 
\begin{align}
& d_{1z} = \frac{\omega\Delta \delta u^{(+)}_{2}}{(\omega^2-2\Delta^2)+\frac{1}{2}\lambda F^s_2(\omega^2-4\Delta^2)}\\
& d_{-1z}' =\frac{\omega\Delta \delta u^{(+)}_{-2}}{(\omega^2-2\Delta^2)+\frac{1}{2}\lambda F^s_2(\omega^2-4\Delta^2)}. 
\end{align}
\end{subequations}
Finally we look at the transverse fluctuation by looking at the $x$ component. 
\begin{subequations}
\begin{align}
& (\omega^2-2\Delta^2)d_{1x} = -i\omega\Delta (\hat{\mb z}\times\delta\bm{\varepsilon}^{(+)}_2)_x\\
& (\omega^2-2\Delta^2)d'_{-1x} = -i\omega\Delta(\hat{\mb z}\times\delta\bm{\varepsilon}^{(+)}_{-2})_x.
\end{align}
\end{subequations}
The spin-dependent self-energy now takes the form 
\begin{align}
\hat{\mb z}\times \delta\bm{\varepsilon}^{(+)}=\hat{\mb z}\times &\delta\mb u^{(+)}+  \int \frac{d\theta'}{2\pi}F^a(\theta, \theta')\Big[-\lambda \hat{\mb z}\times\delta\bm{\varepsilon}^{(+)}-i\frac{\omega}{2}\bar{\lambda}\hat{\mb z}\times [(\bm{\Delta}^*\times\mb d)+(\bm{\Delta}\times\mb d')]\Big].
\end{align}
Projecting it to $\ell = \pm 2$ allows to solve 
\begin{align}
d_{1x} = \frac{-i\omega\Delta (\hat{\mb z}\times\delta\mb u^{(+)})_2}{(\omega^2-2\Delta^2)+\frac{1}{2}\lambda F_2^a(\omega^2-4\Delta^2)}.
\end{align}
To sum up, the analyses in this section have shown the following: (i) The Goldstone modes are not gapped by short-range interaction parametrized by Landau parameters. (ii) For $p$-wave superconductors in both B-phase and A-phase, the sub-gap modes $\sqrt 2\ \Delta$ receives renormalization from quadrupolar Landau parameters $F_2^s$ or $F_2^a$. (iii) For all chiral ground states of finite orbital momenta $L$, the sub-gap modes parallel to their ground states receive mass renormalization from the channel $F_{2L}^s$. The mass corrections referred to in (ii) and (iii) are all determined by the following equation
\begin{align}\label{MassEqn}
(\omega^2-2\Delta^2) +\frac{1}{2}\lambda(\omega) F(\omega^2-4\Delta^2) =0.
\end{align}  
In figure~\ref{fig1} we plot the numerical solution to~\eqref{MassEqn} as a function of $F$, which stands for the Landau parameter of the channel of interest. In accord with the intuition we acquired from the small $F$ expansion, a strong repulsive interaction in particle-hole channel increases the magnitude of the gap, which asymptotically approaches pair-breaking threshold $2\Delta$. On the other hand, an attractive interaction softens the mass of order parameter. In particular, we see the mode would become massless as $F=-1$, at which Pomeranchuk instability of 2 dimensional Fermi liquid is triggered.

We can then look at the region $F = -1+\epsilon$ with $\epsilon\ll 1$. At $T=0$ we can expand the equation \eqref{MassEqn} around $\omega^2 \approx 0$ and extract its dependence on $F$ near the instability. Using the closed form~\eqref{Tsunedo1}, we can deduce the equation~\eqref{MassEqn} has the zero at
\begin{align}
\omega^2 = \frac{3(1+F)}{6+F}\times 4\Delta^2 \approx \frac{12}{5}\epsilon\Delta^2.
\end{align} 
This expression allows us to study how this mode becomes massless as we approaches the instability. 

\section{Field Theory Model for Chiral $p$-wave superfluid}\label{FT}

A complete kinetic theory treatment for both spin-singlet and spin-triplet chiral superfluids in the previous section has been conducted. However, it is still tempting to acquire an effective theory formulation, which would allow us to investigate the problem with techniques and insights across communities. Here, we propose a toy microscopic model for $L = 1$ $p$-wave chiral superfluid at $T=0$. The corresponding Pomeranchuk instability in $2L = 2$ channel triggers the charge nematic order. In the approximation consistent with the kinetic theory approach, it reproduces exactly the same result, and moreover reveals the spin-$2$ nature of the subgap modes of interest.  

Let us consider a 2 dimensional spin-polarized non-relativistic fermion $\psi$ with the kinetic term $ H_{\rm F}[\psi]$, a short range pairing potential in $L=1$ channel $H_{V}$ \cite{Volovik1988}, and a quadrupolar density interaction $H_Q$ devised in Ref. \cite{PhysRevB.64.195109}. Note that the spin degree of freedom is frozen in this regard, and therefore we no longer have the SO${}_S$(3) symmetry to start with. The model thus describes a minimalistic $p$-wave superfluid. The system presented in the previous section in this sense is considered to be 3 copies of the model here. Nonetheless, these ingredients suffice to produce the sub-gap modes and their renormalization. The exact form of the models read 
\begin{subequations}
\begin{align}
\label{HF}& H_{\rm F} = \int \frac{\d^2k}{(2\pi)^2}\, \psi^{\dag}_{\mb k}\bigg(\frac{\mb k^2}{2m}-\epsilon_F\bigg)\psi_{\mb k}:= \int \frac{\d^2\mb k}{(2\pi)^2}\psi^{\dag}_{\mb k}\xi_{\mb k}\psi_{\mb k}\\
\label{HV}& H_V = -\frac{1}{p_F^2}\int \frac{\d^2k}{(2\pi)^2}\frac{\d^2k'}{(2\pi)^2}V_1\psi^{\dag}_{-\mb k}\psi^{\dag}_{\mb k}{\mb k}\cdot{\mb k}'\psi_{\mb k'}\psi_{-\mb k'}\\
\label{HQ}& H_Q = \frac{1}{2}\int \frac{\d^2\mb q}{(2\pi)^2}f_2(\mb q)\mb M(-\mb q)\cdot \mb M(\mb q),
\end{align}
\end{subequations}
where $\mb M = (M_1, M_2)^T$ is defined by the quadrupole moment of particle density
\begin{align}
\begin{pmatrix} M_1 & M_2 \\ M_2 & -M_1\end{pmatrix} = -\frac{1}{p_F^2}\psi^{\dag}\begin{pmatrix} k_x^2-k_y^2 & 2k_xk_y \\ 2k_xk_y & k_y^2-k_x^2\end{pmatrix}\psi
\end{align}
and 
\begin{align}
f_2(\mb q) = \frac{f_2}{1+\kappa\mb q^2} = \nu^{-1}_{\rm 2D}\frac{F_2}{1+\kappa\mb q^2}.
\end{align}
$\nu_{\rm 2D} = \frac{m}{2\pi}$, denoting the density of state of 2D electron gas, and $F_2$ is the conventional Landau parameter. Owing to the frozen spin-degree of freedom, the $F_2$ here corresponds to the spin-independent Landau parameter, $F_2^s$ in the previous section. $\kappa$ characterizes the interaction range and is irrelevant in the long wavelength properties explored below. 
We introduce Hubbard-Stratonovich fields $(\phi, \phi^{\dag})$, $(\Phi, \Phi^{\dag})$ and $(\bar Q, Q)$ to decouple the two-body terms and rewrite the full action as 
\begin{align}
\mathcal S = \int \d^3x\, \Psi^{\dag}(i\p_t-H)_{\rm BdG}\Psi -\frac{1}{2V_1}\int \d^3x\, (\phi^{\dag}\phi+\Phi^{\dag}\Phi)+\int \d^3x \bar Q\, f^{-1}_2(-i\nabla)Q,
\end{align}
where $\Psi = \frac{1}{\sqrt 2}(\psi, \psi^{\dag})^T$ and 
\begin{align}
(i\p_t-H)_{\rm BdG} = & \begin{pmatrix} i\p_t -\xi_{(-i\nabla)} & 0 \\ 0 & i\p_t+\xi_{(-i\nabla)}\end{pmatrix}\notag\\
&+p_F^{-2}\begin{pmatrix} Q(\p_x-i\p_y)^2+\bar Q(\p_x+i\p_y)^2 & 0 \\ 0 & -[(\p_x-i\p_y)^2Q+(\p_x+i\p_y)^2\bar Q]\end{pmatrix} \notag\\
& -\frac{i}{p_F}\begin{pmatrix} 0 & \phi (\p_x-i\p_y)+\Phi(\p_x+i\p_y) \\ \phi^{\dag}(\p_x+i\p_y)+\Phi^{\dag}(\p_x-i\p_y) & 0 \end{pmatrix}.
\end{align}
The derivatives are understood to act on all quantities on their right. From the structure of the action, we see $\phi$ and $\Phi$ represent the $p-ip$ and $p+ip$ pairing amplitude respectively. In the mean field limit,
\begin{subequations}
\begin{align}
\label{GS}& p-ip: \lan \phi\ran = \Delta, \lan\Phi\ran = 0\\
& p+ip: \lan \phi \ran = 0, \lan\Phi\ran  = \Delta.
\end{align}
\end{subequations}
On the other hand, $Q$ and $\bar Q$ represent the nematic order parameter. 

To proceed, we consider a ground state with a gapped fermion spectrum from either of the above choices, integrate the fermion sector and compute the bosonic Gaussian fluctuation. In the explicit computation following, we choose the $p-ip$ ground state~\eqref{GS} and shift $\phi\to \Delta+\phi$ so that $\phi$ presents purely the fluctuation. In this scenario, $\phi$ and $\Phi$ would be playing the role of $\mb d\cdot\hat{\mb z}$ in our kinetic approach. Note that the particle-hole symmetry is usually assumed in kinetic theory, whereas it is exact only on the Fermi surface. The effective field theory respecting this symmetry would acquire an emergent relativistic covariant form \cite{doi:10.1146/annurev-conmatphys-031214-014350}, even though the microscopic origin might rather respect Galilean symmetry. In long wavelength limit $\mb q\to 0$, the particle-hole symmetry can be implemented by evaluating loop momentum on the Fermi surface and extending the depth of the Fermi sea to infinity. Using the method and Feynman rules summarized in appendix \ref{OneLoop}, the resulting effective action has the form 
\begin{align}
\mathcal S_{\rm eff} = \mathcal S_0[\Delta_0, \phi, \phi^{\dag}]+\int \frac{\d^3q}{(2\pi)^3}\, \bigg( & \bar Q(q)M_{\bar QQ}(q)Q(q)+\Phi^{\dag}(q)M_{\Phi^{\dag}\Phi}(q)\Phi(q)\notag\\
& +M_{Q\Phi}(q)Q(-q)\Phi(q) +M_{\bar Q\Phi^{\dag}}(q)\bar Q(-q)\Phi^{\dag}(q)\bigg).
\end{align}
The leading part $\mathcal S_0$ contains the mean field free energy, Goldstone fluctuations $(\phi^{\dag}-\phi)/i$ and the amplitude mode $\phi^{\dag}+\phi$. The bare masses of $Q$ and $\Phi$ are given by the zeros of $M_{\bar QQ}$ and $M_{\Phi^{\dag}\Phi}$, whose explicit forms are given by 
\begin{align}
& M_{\bar QQ} = \nu_{\rm 2D} \bigg(\lambda(\omega)+\frac{1}{F_2}\bigg)\\
& M_{\Phi^{\dag}\Phi} = \frac{\nu_{\rm 2D}\lambda(\omega)}{8\Delta^2}(\omega^2-2\Delta^2).
\end{align}
$\lambda$ is again the Tsunedo function~\eqref{Tsunedo1}. Hence, the bare mass of $Q$ depends on the parameter $F_2$ and becomes soft as $F_2\to -1$. On the other hand, the mass of $\Phi$ is shown to be $m_{\Phi} = \sqrt{2}\, \Delta$, in agreement with the result~\eqref{chiralp1} and~\eqref{chiralp2}. The Fermi-liquid correction can now be understood in terms of the coupling $Q\Phi$ and $\bar Q\Phi^{\dag}$  
\begin{subequations}
\begin{align}
M_{Q\Phi} = M_{\bar Q\Phi^{\dag}} = -\frac{\nu_{\rm 2D}\omega\lambda(\omega)}{4\Delta},
\end{align}
\end{subequations}
indicating $\Phi$ and $Q$ are actually not independent modes. As $F_2 \neq -1$ and $Q$ has a finite bare mass, we are able to integrate out $Q$ to obtain a more compact effective theory. 
\begin{align}
\mathcal S_{\rm eff} = \mathcal S_0 +\int \frac{\d^3q}{(2\pi)^3}\, \frac{\nu_{\rm 2D}\bar{\lambda}(\omega)}{8(1+F_2\lambda(\omega))}\Phi^{\dag}(\omega)\Big[(\omega^2-2\Delta^2)+\frac{1}{2}\lambda(\omega)F_2(\omega^2-4\Delta^2)\Big]\Phi(\omega),
\end{align}
reproducing explicitly the result~\eqref{MassEqn}. Alternatively, one could integrate out $\Phi$ to derive an effective theory of $Q$. 
\begin{align}
\mathcal S_{\rm eff} = \mathcal S_0 +\int \frac{\d^3q}{(2\pi)^3}\nu_{\rm 2D}\bar Q(\omega)\bigg[\frac{\omega^2-4\Delta^2}{2(\omega^2-2\Delta^2)}\lambda(\omega)+\frac{1}{F_2}\bigg] Q(\omega).
\end{align}
Straightforward investigation shows the renormalized mass of $Q$ in the above action is still given by~\eqref{MassEqn}. 

In addition to reproducing the known result, the field theory approach already offers some implications beyond the semi-classical kinetic theory approach:
\begin{enumerate}
\item The exact value $\sqrt 2\, \Delta$ is closely related to the assumption of particle-hole symmetry, or the approximate relativistic nature of the fermionic superfluid on the Fermi surface. Loosening this approximation allows corrections of order $\Delta/\epsilon_F$. Moreover, a term $\Phi^{\dag} i\p_t\Phi$ appears in the action if we breaks particle-hole symmetry during computation, which in turn modifies the value of the bare mass $m_{\Phi}$ as well. In this computation we impose the particle-hole symmetry in order to be consistent with the assumptions of the kinetic theory. While one could compute non-universal corrections to the value of $m_{\Phi}$ by breaking the particle-hole symmetry, we comment that in the weak-coupling computation terms odd in frequency merely change the mass slightly and the magnitude of $m_{\Phi}$ would remain $\mathcal O(\Delta)$. The qualitative fact that this mass is reduced by negative $F_2$ is not affected.
\item In the presence of a condensate, operators are classified by the residual symmetry respected by the ground state. Taking the $p-ip$ ground state for instance, 
\begin{align}
\Delta_{\mb p} \sim (p_x-ip_y)\lan \psi(-\mb p)\psi(\mb p)\ran.
\end{align}
$\Delta_{\mb p}$ is symmetric under a combination of U(1) charge transformation and orbital rotation 
\begin{subequations}
\begin{align}
 \psi & \to e^{i\alpha/2}\psi\\
 \begin{pmatrix}p_x\\ p_y \end{pmatrix} & \to \begin{pmatrix}p_x\cos\alpha-p_y\sin\alpha\\
  p_y\cos\alpha+p_x\sin\alpha\end{pmatrix}.
\end{align}
\end{subequations}
In this example, the operators are classified using the {\it angular momentum $\ell$} defined by this combined transformation $\mathcal O\to e^{i\ell\alpha}\mathcal O$. In particular, the fluctuation of $p+ip$ condensate transforms as
\begin{align}
(p_x+ip_y)\lan \psi(-\mb p)\psi(\mb p)\ran \to e^{2i\alpha} (p_x+ip_y)\lan \psi(-\mb p)\psi(\mb p)\ran
\end{align}
and has angular momentum $\ell = $2. Similarly, the nematic order parameter transforms as
\begin{align}
(p_x+ip_y)^2\lan \psi^{\dag}(\mb p)\psi(\mb p)\ran \to e^{2i\alpha}(p_x+ip_y)^2\lan \psi^{\dag}(\mb p)\psi(\mb p)\ran,
\end{align}
indicating both operators possess the {\it spin-2} nature under the residual symmetry group. The {\it spin-2$L$} states can also be understood from this perspective. Besides, that $\Phi$ and $Q$ are not independently fluctuating can be explained in terms of the notion of emergent geometry \cite{PhysRevLett.119.146602, PhysRevB.98.064503}. The nematic order parameters $Q:= (Q_1+iQ_2)/2$ and $\bar Q:= (Q_1-iQ_2)/2$, under a proper normalization \footnote{In the present work, we have to replace $Q\to -\epsilon_FQ$.}, also parametrize an emergent unimodular metric $\mathfrak g_{ij}$ via 
\begin{align}
\mathfrak g:= \exp\begin{pmatrix} Q_1 & Q_2 \\ Q_2 & -Q_1\end{pmatrix}.
\end{align}
Similarly, the order parameters of a $p$-wave superfluid $\Delta^i$ also define an emergent geometric degree of freedom $\mathfrak G^{ij}\sim \Delta^{*i}\Delta^j$. The sub-gap modes, in this language, correspond to the spin-2 sector of $\mathfrak G$. Placed on a flat space and close to equilibrium, both $\mathfrak g$ and $\mathfrak G$ favor the Euclidean metric $\delta_{ij}$. Hence, their fluctuations, both being spin-2, are indistinguishable from this geometric perspective. 
\item In previous studies these modes are usually overlooked regarding low energy physics \cite{PhysRevB.89.174507}, even though they are responsible for electromagnetic response at high frequency \cite{PhysRevB.62.3042}. That the spin-2 mode becomes soft as $F_2= -1$ suggests that an effective low energy theory different from one in Ref. \cite{PhysRevB.89.174507} should be formulated to incorporate a spin-2 mode close to a nematic critical point. While different microscopic models could produce different results depending on model dependent parameters, symmetry principle together with the above geometric picture suggests an effective action that replaces the background geometry with the internal fluctuating geometry. A proper microscopic model is then responsible for correctly producing effects such as the analogous Hall viscosity, which comes from a $\bar Q\p_tQ$ term in effective action. The existence of this term breaks time reversal symmetry and distinguish a $p-ip$ ground state from a $p+ip$ one. This issue is beyond the scope of current work and will be addressed in detail separately in the future.
\end{enumerate}
\section{Conclusion}

In conclusion, we revisit a class of 2 dimensional superfluids. Using the semi-classical kinetic equation, we compute the order parameter collective modes for 2 dimensional B-phase and general chiral ground states of angular momenta $L\geq 0$. Extending the known results for $L=1$, we show that the sub-gap modes of the universal mass value $\sqrt{2}\, \Delta$ exist for all chiral ground states $L\geq 1$ in the limit with exact particle-hole symmetry. By renormalizing the fermionic self-energy, we calculate the correction of these sub-gap modes from Fermi-liquid corrections and discover those sub-gap modes, sourced by $F_{2L}$, could be regarded as spin-$2L$ modes, where $L$ is the angular momentum of their underlying ground state. The masses increase for repulsive fermionic interactions and soften for attractive ones. Remarkably, renormalized sub-gap modes become gapless when the Pomeranchuk instability in the corresponding channel is triggered. 

Moreover, we proposed a toy model for the case $L=1$, whose effective bosonic action is able to reproduce the kinetic result under the consistent approximations. This model could describe a $p$-wave chiral superfluid near a nematic critical point, and furthermore allows us to loosen the common assumptions made in semi-classical approaches and utilize the insights from field theory communities to understand the nature of the sub-gap modes. The author hopes the approaches and conclusions drawn from this work could provide the studies of quantum Hall nematic physics and nematic unconventional superfluid a complementary perspective and new insights.

\acknowledgements
The author thanks O. Golan, E. Berg, K. Levin and J. A. Sauls for valuable suggestions, and grateful for Dam Thanh Son, Yu-Ping Lin and Chien-Te Wu for comments on the manuscript. This work is supported, in part, by U.S.\ DOE grant
No.\ DE-FG02-13ER41958
and a Simons Investigator Grant from the Simons Foundation.  Additional
support was provided by the Chicago MRSEC, which is funded by NSF
through grant DMR-1420709. 
\appendix 

\section{$\gamma$ and the Tsunedo function $\lambda$}\label{Fxns}
The integral $\gamma$ is 
\begin{align}\label{gammafunction}
\int_{|\Delta|}^{\infty}\frac{d\varepsilon}{2\pi i}\frac{1}{\sqrt{\varepsilon^2-|\Delta|^2}}\tanh\frac{\varepsilon}{2T} = \gamma.
\end{align}
It is formally divergent, but can be regularized and identified with $1/V_{\ell}$ (or $1/(2V_0)$) using linearized gap equation. 

The function $\lambda$ was first introduced by Tsunedo as a kind of Cooper pair susceptibility. 
\begin{align}
&\bar{\lambda}(\hat{\mb p};\omega,q ) = \frac{\lambda}{|\Delta(\hat{\mb p})|^2}\notag\\
 =&\int^{\infty}_{-\infty}\frac{d\varepsilon}{2\pi i}\frac{\mathsf n(\varepsilon_-)(2\varepsilon\omega-\eta^2)-\mathsf n(\varepsilon_+)(2\varepsilon\omega+\eta^2)}{(4\varepsilon^2-\eta^2)(\omega^2-\eta^2)+4|\Delta|^2\eta^2},
\end{align}
where $\eta = v_F\mb q\cdot\hat{\mb p}$ and
\begin{align}\label{Littlen}
\mathsf n(\varepsilon) = -\frac{2\pi i\mathrm{sgn}(\varepsilon)}{\sqrt{\varepsilon^2-|\Delta|^2}}\Theta(\varepsilon^2-|\Delta|^2)\tanh\frac{\varepsilon}{2T}.
\end{align}
In $q\to 0$ limit, the integral reduces to 
\begin{align}
\lambda = |\Delta|^2\int_{|\Delta|}^{\infty}\frac{d\varepsilon}{\sqrt{\varepsilon^2-|\Delta|^2}}\frac{\tanh\frac{\varepsilon}{2T}}{\varepsilon^2-\omega^2/4}.
\end{align}
These expressions can be used in numerical evaluation. This function actually has an analytic closed form in the limit $T\to0$. Writing $x = \omega/(2|\Delta|)$, 
\begin{subequations}
\begin{align}
\label{Tsunedo1}\lambda(\omega) = \frac{\sin^{-1}x}{x\sqrt{1-x^2}},\ |x|<1,
\end{align}
where as for $|x|>1$,
\begin{align}
\label{Tsunedo2}\lambda(\omega) =\textstyle \frac{1}{2x\sqrt{x^2-1}}\Big[\log\Big|\frac{\sqrt{x^2-1}-x}{\sqrt{x^2-1}+x}\Big|+i\pi \mathrm{sgn}(x)\Big].
\end{align}
\end{subequations}

\section{Full Dynamical Equations}\label{FDE}
In this section we sketch the steps for inverting the kinetic equation and give the full dynamical equations at finite wavelength. Expanding~\eqref{EoM} with respect to the ground state of interest, we could found components of the Keldysh Green's function satisfy a general equation  
\begin{align}\label{MEqn}
\bm{\Omega} |\widehat g\ran = \mathsf M|\widehat{\sigma}\ran.
\end{align}

The quasi-classical Green's functions can thus be obtained via
\begin{align}
\int\frac{d\varepsilon}{2\pi i}|\widehat{g}\ran = \int \frac{d\varepsilon}{2\pi i}\bm{\Omega}^{-1}\mathsf M|\widehat{\sigma}\ran.
\end{align}
We note that when performing $\varepsilon$ integral in this work, the particle-hole symmetry $\varepsilon\leftrightarrow -\varepsilon$ is assumed. 

The defined in~\eqref{MEqn} the matrices are 
\begin{align}
\bm{\Omega} = \begin{bmatrix} -\eta & \omega & 2i\Delta_I & -2\Delta_R\\ \omega & -\eta & 0 & 0 \\ 2i\Delta_I & 0 & -\eta & 2\varepsilon \\ 2\Delta_R & 0 & 2\varepsilon & -\eta\end{bmatrix}
\end{align}
and 
\begin{align}
\mathsf M =\begin{bmatrix}
0 & -\mathsf m_a & -i\mathsf n_s\Delta_I & \mathsf n_s\Delta_R\\
-\mathsf m_a & 0 & -\mathsf n_a\Delta_R & i\mathsf n_a\Delta_I\\
-i\Delta_I\mathsf n_s & \Delta_R\mathsf n_a & 0 & -\mathsf m_s\\
-\mathsf n_s\Delta_R & i\mathsf n_a\Delta_I & -\mathsf m_s & 0
\end{bmatrix}.
\end{align}
$\Delta_R$ and $\Delta_I$ are the real and imaginary parts of the gap function. In terms of the $\mathsf n$ defined by~\eqref{Littlen}, the elements in $\mathsf M$ are 
\begin{subequations}
\begin{align}
& \mathsf n_s = \mathsf n(\varepsilon_+)+\mathsf n(\varepsilon_-)\\
& \mathsf n_a = \mathsf n(\varepsilon_+)-\mathsf n(\varepsilon_-)\\
& \mathsf m = \varepsilon\mathsf n(\varepsilon)\\
& \mathsf m_s =\mathsf m(\varepsilon_+)+\mathsf m(\varepsilon_-) \\
& \mathsf m_a = \mathsf m(\varepsilon_+)-\mathsf m(\varepsilon_-).
\end{align}
\end{subequations}

In the rest of the section we give the proper combinations $|\widehat{g}\ran$ and $|\widehat{\sigma}\ran$ and the complete dynamical equations.

\subsection{Singlet-pairing ground state}
For a singlet-pairing state, the bosonic fluctuation couples only to spin independent fermionic self-energies, and the relevant equations are those which $\delta g$, $\delta g'$, $d$ and $d'$ obey. These equations can be easily solved by taking 
\begin{align}
|\widehat{g}\ran = \begin{pmatrix} \delta g^{(-)} \\ \delta g^{(+)} \\ \delta f^{(+)} \\ \delta f^{(-)}\end{pmatrix}, |\widehat{\sigma}\ran = \begin{pmatrix} \delta\varepsilon^{(-)} \\ \delta\varepsilon^{(+)} \\ d^{(+)} \\ d^{(-)}\end{pmatrix}.
\end{align}
Expressing $|\widehat{g}\ran$ in terms of $|\widehat{\sigma}\ran$ and performing convolutions with suitable potentials would imply the following equations.
\begin{widetext}
\begin{subequations}

\begin{align}
\delta\varepsilon^{(-)}(\hat{\mb p}; \omega, \mb q) -\delta\varepsilon^{(+)}_{\rm ext}= \int \frac{d\theta'}{2\pi} &A^s(\theta, \theta')\Big\{\Big(1+(1-\lambda(\hat{\mb p}'))\frac{{\eta'}^2}{\omega^2-{\eta'}^2}\Big)\delta\varepsilon^{(-)}(\hat{\mb p}')\notag\\
& +\frac{\omega\eta'}{\omega^2-{\eta'}^2}(1-\lambda(\hat{\mb p}'))\delta\varepsilon^{(+)}(\hat{\mb p}')+\frac{\bar{\lambda}(\hat{\mb p}')\eta'}{2}[d(\hat{\mb p}')\Delta^*(\hat{\mb p}')-\Delta(\hat{\mb p}') d(\hat{\mb p}')] \Big\}.
\end{align}

\begin{align}
\delta\varepsilon^{(+)}(\hat{\mb p}; \omega,\mb q) -\delta\varepsilon^{(+)}_{\rm ext}= \int \frac{d\theta'}{2\pi}&A^s(\theta, \theta')\Big\{ \frac{\omega\eta'}{\omega^2-{\eta'}^2}(1-\lambda(\hat{\mb p}'))\delta\varepsilon^{(-)}(\hat{\mb p}')\notag\\
&+\frac{\omega^2}{\omega^2-{\eta'}^2}(1-\lambda(\hat{\mb p}'))\delta\varepsilon^{(+)}(\hat{\mb p}')-\frac{1}{2}\omega\bar{\lambda}(\hat{\mb p}')[d'(\hat{\mb p}')\Delta(\hat{\mb p}')-d(\hat{\mb p}')\Delta^*(\hat{\mb p}')]\Big\}.
\end{align}
\begin{align}
d(\hat{\mb p}; \omega, \mb q) = \int \frac{d\theta'}{2\pi}V_{\rm e}(\theta, \theta') & \Big\{\Big(\gamma+\frac{1}{4}\bar{\lambda}(\hat{\mb p}')(\omega^2-{\eta'}^2-2|\Delta(\hat{\mb p}')|^2)\Big)d(\hat{\mb p}')\notag\\
&-\frac{\bar{\lambda}(\hat{\mb p}')}{2}\Delta^2(\hat{\mb p}')d'(\hat{\mb p}')-\frac{\Delta(\hat{\mb p}')\bar{\lambda}(\hat{\mb p}')}{4}(\eta'\delta\varepsilon^{(-)}(\hat{\mb p'})+\omega\delta\varepsilon^{(+)}(\hat{\mb p'})]\Big\}.
\end{align}

\begin{align}
d'(\hat{\mb p}; \omega, \mb q) = \int \frac{d\theta'}{2\pi}V_{\rm e}(\theta, \theta')&\Big\{\Big(\gamma+\frac{1}{4}\bar{\lambda}(\hat{\mb p}')(\omega^2-{\eta'}^2-2|\Delta(\hat{\mb p}')|^2)\Big)d'(\hat{\mb p}')\notag\\
&-\frac{\bar{\lambda}(\hat{\mb p}')}{2}(\Delta^*(\hat{\mb p}'))^2d(\hat{\mb p}')+\frac{\Delta^*(\hat{\mb p}')\bar{\lambda}(\hat{\mb p}')}{4}[\eta'\delta\varepsilon^{(-)}(\hat{\mb p}')+\omega\delta\varepsilon^{(+)}(\hat{\mb p}')]\Big\}.
\end{align}

\end{subequations}
\end{widetext}
\subsection{Triplet-pairing ground state}
For a triplet pairing ground state, the vector $\mb d$ and $\mb d'$ couple to both spin-dependent and independent self-energies. We denote the direction of ground state condensate $\bm{\Delta}$ as $\hat{\mb n}$. To solve vector quantities $\mb d$ and $\bm{\varepsilon}$, we decompose them into components longitudinal $L$ and transverse $T$ to the gap $\hat{\mb n}$, that is, a vector $\mb v$ is decomposed as $\mb v =\mb v_L+\mb v_T$, where $\mb v_L = \hat{\mb n}(\mb v\cdot\hat{\mb n})$. The complete set of equations can be solved by considering the following combinations of $\{|\widehat{g}\ran, |\widehat{\sigma}\ran\}$.
The part coupled with spin-independent $\delta g$ is the longitudinal modes
\begin{subequations}
\begin{align}
\Big\{\begin{pmatrix} \delta g^{(-)}\\ \delta g^{(+)} \\ \delta\mb f_L^{(+)} \\ \delta \mb f_L^{(-)}\end{pmatrix},\begin{pmatrix} \delta\varepsilon^{(-)}\\ \delta\varepsilon^{(+)} \\ \mb d^{(+)}_L \\ \mb d^{(-)}_L\end{pmatrix}\Big\}.
\end{align}
The part coupled with spin-dependent $\delta\mb g$, on the other hand, includes the transverse and binormal parts of the anomalous Green's function.   
\begin{align}
\Big\{\begin{pmatrix} \hat{\mb n}\times\delta \mb g^{(-)}\\ \hat{\mb n}\times\delta \mb g^{(+)} \\ i\delta\mb f_T^{(-)} \\ i\delta \mb f_T^{(+)}\end{pmatrix},\begin{pmatrix} \hat{\mb n}\times\delta\bm{\varepsilon}^{(-)}\\ \hat{\mb n}\times\delta\bm{\varepsilon}^{(+)} \\ i\mb d^{(-)}_T \\ i\mb d^{(+)}_T\end{pmatrix}\Big\}.
\end{align}
\begin{align}
\Big\{\begin{pmatrix} \delta\mb g^{(-)}_T\\ \delta \mb g^{(+)}_T \\ i\delta\mb f^{(-)}\times\hat{\mb n} \\ i\delta \mb f^{(+)}\times\hat{\mb n}\end{pmatrix},\begin{pmatrix} \delta\bm{\varepsilon}^{(-)}_T\\ \delta\bm{\varepsilon}^{(+)}_T \\ i\mb d^{(-)}\times\hat{\mb n} \\ i\mb d^{(+)}\times\hat{\mb n}\end{pmatrix}\Big\}.
\end{align}
The above 2 sets of vectors give only the binormal and transverse information about $\delta\mb g.$  It turns out the spin-singlet components $\delta f$ and $d$ are required to access the longitudinal information of $\delta\mb g$ using the combination below. 

\begin{align}
\Big\{\begin{pmatrix} \delta \mb g^{(+)}_L\\ \delta \mb g^{(-)}_L \\ \delta f^{(+)} \\ \delta f^{(-)}\end{pmatrix},\begin{pmatrix} \delta\bm{\varepsilon}^{(+)}_L\\ \delta\bm{\varepsilon}^{(-)}_L \\  d^{(+)} \\  d^{(-)}\end{pmatrix}\Big\}.
\end{align}
These spin-singlet degrees of freedom $\delta f$ and $d$ are treated as external sources and turned off at the end of computation. After solving all above $|\widehat{g}\ran$ in terms of $|\widehat{\sigma}\ran$, we could again make use of~\eqref{F1},~\eqref{F2},~\eqref{G1}, and~\eqref{G2} to obtain the following equations.
\end{subequations}
\begin{widetext}
\begin{subequations}
\begin{align}
\delta\varepsilon^{(-)}(\hat{\mb p}; \omega, \mb q)-\delta\varepsilon^{(-)}_{\rm ext} = \int& \frac{d\theta}{2\pi}  \, A^s(\theta, \theta')\Big\{\Big(1+(1-\lambda(\hat{\mb p}'))\frac{{\eta'}^2}{\omega^2-{\eta'}^2}\Big)\delta\varepsilon^{(-)}(\hat{\mb p}')\notag\\
&+\frac{\omega{\eta'}}{\omega^2-{\eta'}^2}(1-\lambda(\hat{\mb p}'))\delta\varepsilon^{(+)}(\hat{\mb p}')+\frac{1}{2}{\eta'}\bar{\lambda}(\hat{\mb p}')[\bm{\Delta}^*(\hat{\mb p}')\cdot\mb{d}(\hat{\mb p}')-\bm{\Delta}(\hat{\mb p}')\cdot\mb{d}'(\hat{\mb p}')]\Big\}.
\end{align}
\begin{align}
\delta \varepsilon^{(+)}(\hat{\mb p}; \omega, \mb q)-\delta\varepsilon^{(+)}_{\rm ext} = \int &\frac{d\theta'}{2\pi}\,  A^s(\theta, \theta')\Big\{ \frac{\omega{\eta'}}{\omega^2-{\eta'}^2}(1-\lambda(\hat{\mb p}'))\delta\varepsilon^{(-)}(\hat{\mb p}')\notag\\
& +\frac{\omega^2}{\omega^2-{\eta'}^2}(1-\lambda(\hat{\mb p}'))\delta\varepsilon^{(+)}(\hat{\mb p}')+\frac{1}{2}\omega\bar{\lambda}(\hat{\mb p}')[\bm{\Delta}^*(\hat{\mb p}')\cdot\mb d(\hat{\mb p}')-\bm{\Delta}(\hat{\mb p}')\cdot\mb{d}'(\hat{\mb p}')]\Big\}.
\end{align}

\begin{align}
\delta\bm{\varepsilon}^{(-)}(\hat{\mb p};\omega, \mb q) = \int \frac{d\theta'}{2\pi} & A^a(\theta, \theta')\Big\{\Big(1+(1-\lambda(\hat{\mb p}'))\frac{{\eta'}^2}{\omega^2-{\eta'}^2}\Big)\delta\bm{\varepsilon}^{(-)}(\hat{\mb p}')+\frac{\omega\eta'}{\omega^2-{\eta'}^2}(1-\lambda(\hat{\mb p}'))\delta\varepsilon^{(+)}(\hat{\mb p}')\notag\\
&-\lambda(\hat{\mb p}')[\delta\bm{\varepsilon}^{(-)}(\hat{\mb p}')\cdot\hat{\mb n}(\hat{\mb p}')]\hat{\mb n}(\hat{\mb p}')-\frac{i\eta'\bar{\lambda}(\hat{\mb p}')}{2}[\bm{\Delta}^*(\hat{\mb p}')\times\mb d(\hat{\mb p}')+\bm{\Delta}(\hat{\mb p}')\times\mb d'(\hat{\mb p}')]\Big\}.
\end{align}

\begin{align}
\delta\bm{\varepsilon}^{(+)}(\hat{\mb p}; \omega, \mb q) = \int \frac{d\theta'}{2\pi}& A^a(\theta, \theta')\Big\{\frac{\omega^2}{\omega^2-{\eta'}^2}(1-\lambda(\hat{\mb p'}))\delta\bm{\varepsilon}^{(+)}(\hat{\mb p}')+\frac{{\eta'}\omega}{\omega^2-{\eta'}^2}(1-\lambda(\hat{\mb p}'))\delta\bm{\varepsilon}^{(-)}(\hat{\mb p}')\notag\\
& +\lambda(\hat{\mb p}')[\delta\bm{\varepsilon}^{(+)}(\hat{\mb p}')\cdot\hat{\mb n}(\hat{\mb p}')]\hat{\mb n}(\hat{\mb p}')-\frac{i\omega}{2}\bar{\lambda}(\hat{\mb p}')[\bm{\Delta}^*(\hat{\mb p}')\times\mb d(\hat{\mb p}')+\bm{\Delta}(\hat{\mb p}')\times\mb d'(\hat{\mb p}')]\Big\}.
\end{align}

\begin{align}\label{vd3}
\mb d(\hat{\mb p}; \omega, \mb q) =& \int \frac{d\theta'}{2\pi}V_{\rm o}(\theta, \theta')\Big\{\Big(\gamma+\frac{1}{4}\bar{\lambda}(\hat{\mb p}')(\omega^2-{\eta'}^2-2|\Delta(\hat{\mb p}')|^2)\Big)\mb d(\hat{\mb p}')\notag\\
&-\frac{1}{4}\bar{\lambda}(\hat{\mb p}')\bm{\Delta}(\hat{\mb p}')[{\eta'}\delta\varepsilon^{(-)}(\hat{\mb p}')+\omega\delta\varepsilon^{(+)}(\hat{\mb p}')]+\frac{1}{4}\bar{\lambda}(\hat{\mb p}')i\bm{\Delta}(\hat{\mb p}')\times({\eta'}\delta\bm{\varepsilon}^{(-)}(\hat{\mb p}')+\omega\delta\bm{\varepsilon}^{(+)}(\hat{\mb p}'))\notag\\
&\ \ \ \ \ \ \ \ \ \ \ \ \ \ \ \ \ \ \ \ \ \ \ \ \ \ \ \ \ \ \ \ \ \ \ \ \ \ \ \ \ \ +\frac{1}{2}\bar{\lambda}(\hat{\mb p}')[(\bm{\Delta}(\hat{\mb p}')\cdot\bm{\Delta}(\hat{\mb p}'))\mb d'(\hat{\mb p}')-2(\bm{\Delta}(\hat{\mb p}')\cdot\mb d'(\hat{\mb p}'))\bm{\Delta}(\hat{\mb p}')]\Big\}.
\end{align}
\begin{align}\label{vd4}
\mb d'(\hat{\mb p}; \omega, \mb q) =& \int \frac{d\theta'}{2\pi}V_{\rm o}(\theta, \theta')\Big\{\Big(\gamma+\frac{1}{4}\bar{\lambda}(\hat{\mb p}')(\omega^2-{\eta'}^2-2|\Delta(\hat{\mb p}')|^2)\Big)\mb d'(\hat{\mb p}')\notag\\
&+\frac{1}{4}\bar{\lambda}(\hat{\mb p}')\bm{\Delta}^*(\hat{\mb p}')[{\eta'}\delta\varepsilon^{(-)}(\hat{\mb p}')+\omega\delta\varepsilon^{(+)}(\hat{\mb p}')]+\frac{1}{4}\bar{\lambda}(\hat{\mb p}')i\bm{\Delta}^*(\hat{\mb p}')\times({\eta'}\delta\bm{\varepsilon}^{(-)}(\hat{\mb p}')+\omega\delta\bm{\varepsilon}^{(+)}(\hat{\mb p}'))\notag\\
&\ \ \ \ \ \ \ \ \ \ \ \ \ \ \ \ \ \ \ \ \ \ \ \ \ \ \ \ \ \ \ \ \ \ \ \ \ \ \ \ \ \ +\frac{1}{2}\bar{\lambda}(\hat{\mb p}')[(\bm{\Delta}^*(\hat{\mb p}')\cdot\bm{\Delta}^*(\hat{\mb p}'))\mb d(\hat{\mb p}')-2(\bm{\Delta}^*(\hat{\mb p}')\cdot\mb d(\hat{\mb p}'))\bm{\Delta}^*(\hat{\mb p}')]\Big\}.
\end{align}
\end{subequations}
\end{widetext}

\section{1-loop action computation}\label{OneLoop}
The system concerning us in this note is a spin polarized  $p$-wave chiral superfluid. We adapt the simplest pairing model induced by a contact pairing, which we can decouple by introducing a dynamical auxiliary field via Hubbard-Stratonovich transformation. We also introduce the nematic fluctuation in particle-hole channel. Those interactions can also be decoupled by introducing more auxiliary/collective fields, which we will denote as $\phi_I$ in the following. 

The fermionic part of the action after possibly multiple Hubbard-Stratonovich transformations can be written as $S = \int (\d x)\, \Psi^{\dag}iD^{-1}\Psi$, where $\Psi$ is the Nambu spinor $\Psi^T = (\psi, \psi^{\dag})$. The partition function of the fermion sector is then
\begin{align}
\int \mathscr D\Psi^{\dag}\mathscr D\Psi \, \exp\bigg(-\int (\d x)\Psi^{\dag}D^{-1}\Psi\bigg):=\exp\bigg(iS_{\rm eff}\bigg).
\end{align}
Since we are only considering a single species of fermions, the effective action of auxiliary fields $\{\phi_I\}$ reads 
\begin{align}
S_{\rm eff} = -\frac{i}{2}\mathrm{Tr}\log D^{-1}.
\end{align}
Formally, we can expand the action with respect to a classical solution $\phi^0_J(k) = \lan \phi^0_J\ran\times(2\pi)^3\delta(k)$. In terms of the deviation $\delta\phi_J = \phi_J-\phi^0_J$
\begin{align}
S =& S_{\rm eff}+S_{\rm aux}[\phi_I] \notag\\
=& S[\phi^0_I]+\int (\d q)\, \frac{\delta S}{\delta\phi_J(q)}[\phi^0]\delta\phi_J(q)+\frac{1}{2}\int (\d q)(\d q')\, \delta\phi_I(q)\frac{\delta^2 S}{\delta\phi_I(q)\delta \phi_J(q')}[\phi^0]\delta\phi_J(q')+\cdots
\end{align}
The first order condition 
\begin{align}
\frac{\delta S}{\delta\phi_I(k)}[\phi^0] = 0
\end{align}
is often used to specify the information of a certain uniform ground state $\phi_I^0(2\pi)^3\delta(k)$. The key ingredient of the Gaussian effective theory is the second derivative of the action evaluated with respect to the ground state. To extract the contribution from $S_{\rm eff}$, we have to compute 
\begin{align}
& \frac{\delta^2}{\delta\phi_I(q)\delta\phi_J(q')}S_{\rm eff} = -\frac{i}{2}\frac{\delta^2}{\delta\phi_I(q)\delta\phi_J(q')}\mathrm{Tr}\log D^{-1}\notag\\
= & -\frac{i}{2}\frac{\delta}{\delta\phi_I(q)}\mathrm{Tr}D\frac{\delta D^{-1}}{\delta\phi_J(q')} = -\frac{i}{2}\mathrm{Tr}\frac{\delta D}{\delta\phi_I(q)}\frac{\delta D^{-1}}{\delta\phi_J(q')}-\frac{i}{2}\mathrm{Tr}D\frac{\delta^2D^{-1}}{\delta\phi_I(q)\delta\phi_J(q')}\notag\\
= & \frac{i}{2}\mathrm{Tr}[D\frac{\delta D^{-1}}{\delta\phi_I(q)}D\frac{\delta D^{-1}}{\delta\phi_J(q')}]-\frac{i}{2}\mathrm{Tr}[D\frac{\delta^2D^{-1}}{\delta\phi_I(q)\delta\phi_J(q')}]
\end{align}
Thus the 2-point function of our interest is then 
\begin{align}\label{two_point_fx}
M^{IJ}(q,q') = \frac{i}{2}\mathrm{Tr}[D\frac{\delta D^{-1}}{\delta\phi_I(q)}D\frac{\delta D^{-1}}{\delta\phi_J(q')}]-\frac{i}{2}\mathrm{Tr}[D\frac{\delta^2D^{-1}}{\delta\phi_I(q)\delta\phi_J(q')}]\bigg|_{\phi=\phi^0}
\end{align}
The second term in~\eqref{two_point_fx} is usually referred to as the contact term. Example includes the diamagnetic current term of electromagnetic response. For the model concerning us in this article, we only have to focus on the first term in~\eqref{two_point_fx}.

To compute the two-point functions presented in the main text, we will need the ground state propagator of a $p-ip$ ground state and the variation of the sum~\eqref{HF}+\eqref{HV}+\eqref{HQ} with respect to fields $\bar Q, Q, \Phi^{\dag}$ and $\Phi$. 
The propagators are 
\begin{subequations}
\begin{align}
& iD^{-1}_0 (p,p')= (2\pi)^3\delta(p-p')\begin{pmatrix} p_0-\xi_{\mb p} & \Delta p_F^{-1}(p_x-ip_y) \\ \Delta p_F^{-1}(p_x+ip_y) & p_0+\xi_{\mb p}\end{pmatrix}\\
& D_0(p,p') = (2\pi)^3\delta(p-p')\frac{i}{p_0^2-E_{\mb p}^2+i\eta}\begin{pmatrix} p_0+\xi_{\mb p} & -\Delta p_F^{-1}(p_x-ip_y) \\ -\Delta p_F^{-1}(p_x+ip_y) & p_0-\xi_{\mb p}\end{pmatrix}
\end{align}
\end{subequations}
with $E_{\mb p}^2 = \xi_{\mb p}^2+p_F^{-2}\Delta^2\mb p^2$. Finally, the vertices are 
\begin{subequations}
\begin{align}
& \frac{\delta iD^{-1}(p,p')}{\delta Q(q)} = \frac{1}{p_F^2}\begin{pmatrix} -(p_x'-ip_y')^2 & 0 \\ 0 & (p_x-ip_y)^2\end{pmatrix} (2\pi)^3\delta(p-p'-q)\\
& \frac{\delta iD^{-1}(p,p')}{\delta \bar Q(q)} = \frac{1}{p_F^2}\begin{pmatrix} -(p_x'+ip_y')^2 & 0 \\ 0 & (p_x+ip_y)^2\end{pmatrix} (2\pi)^3\delta(p'-p-q)\\
& \frac{\delta iD^{-1}(p,p')}{\delta \Phi(q)} = (2\pi)^3\delta(p-p'-q)\begin{pmatrix} 0 & p_F^{-1}(p_x'+ip_y') \\ 0 & 0 \end{pmatrix}\\
& \frac{\delta iD^{-1}(p,p')}{\delta\Phi^{\dag}(q)} = (2\pi)^3\delta(p'-p-q)\begin{pmatrix} 0 & 0 \\ p_F^{-1}(p_x'-ip_y') & 0\end{pmatrix}.
\end{align}

\end{subequations}
\bibliography{citation}{}
\bibliographystyle{apsrev4-1}
\end{document}